\documentclass[11pt,a4paper]{article}
\usepackage{jheppub}
\usepackage{url}
\usepackage[T1]{fontenc}

\newcommand{\beq}{\begin{equation}}
\newcommand{\eeq}{\end{equation}}

\newcommand{\ud}{\mathrm{d}}
\newcommand{\udr}{\mathrm{dir}}

\newcommand{\uim}{\mathrm{im}}

\newcommand{\urv}{\mathrm{v}}
\newcommand{\Ec}{\mathcal{E}}

\newcommand{\Er}{\mathrm{E}}
\newcommand{\Br}{\mathrm{B}}

\title{\boldmath Fields of an  ultra-relativistic beam of charged particles  between parallel plates. \\
Exact 2D solutions  by the method of images and applications to the HL-LHC}

\author{B.~B.~Levchenko}
\affiliation{D.V. Skobeltsyn Institute of Nuclear Physics, M.V. Lomonosov Moscow State University,\\
Moscow 119991, Russian Federation}
\emailAdd{levtchen@mail.desy.de}

\abstract{Exact two-dimensional  analytic expressions for electric and magnetic fields and their potentials created by a linear beam of relativistic charged particles between infinite perfectly conductive plates and ferromagnetic poles are derived.
The solutions are obtained by summing an infinite sequence of fields from linear charge-images and current-images in complex space.
Knowledge of the normal component of the electric field on the conductor surface makes it possible to calculate the induced  electric charge surface density.
In addition, we derive within an improved linear approximation
new  analytical expressions for fields near the beam 
in the case of an arbitrary beam  offset from the median plane.
The mathematical features of exact solutions and limitations for the applicability of linear approximations are specified. 

The primary goals of the future high-luminosity p-p and heavy-ion LHC programme 
after the Long Shutdown 2 
are the search  for yet unobserved effects of physics beyond the Standard Model,  
searches for  rare or low-sensitivity processes in the Higgs sector, and 
probing in more detail the mechanism of electroweak symmetry breaking.
This programme relies on the stable operation of the accelerator.
However, as the beam luminosity increases, a number of destabilizing phenomena occur,
in particular field emission, enhancing the electron cloud effect.
 For the case of a proton beam, we apply the  exact 2D solution for  
estimating the intensity of electron field emission activated by the electric field of 
the beam  in collimators of the future high-luminosity Large Hadron Collider.
Calculation shows that the field emission intensity is very sensitive to a collimator surface roughness.
In addition, with a relatively small and accidental beam displacement from the median path, 
about  20$\%$ of the collimator half-gap, 
the emission intensity increases by a factor of 10$^7$.
This will partially neutralize the beam space charge, violating acceleration dynamics 
and enhancing  instability effects.
}


\begin{document}
%
%
\maketitle
\flushbottom

\section{\label{intro}Introduction}

The external electric field  $\mathbf{E}$ and magnetic induction $\mathbf{B}$ of a beam of relativistic charged particles with a uniform linear density and circular cross section are described by  expressions \cite{Wiedemann:2007yf} (section~18.2.4), \cite{Levchenko:2020}
 %
\beq
 \mathbf{E}_{\rm dir}=\frac{2\kappa\lambda}{r}\frac{\mathbf{r}}{r}\,, \qquad 
 \mathbf{B}_{\rm dir}=\frac{1}{c}\boldmath{\beta}\times \mathbf{E}\,,
\label{I-1}
\eeq
where $\kappa= 1/4\pi\epsilon_0$,  $\lambda$ is  the linear beam density
with account the particle charge,  $\boldsymbol\beta= \mathbf{v}/c$ 
is a normalized velocity vector of the beam constituents and $c$ the velocity of light.
The radius vector $\mathbf{r}$ is perpendicular to the velocity vector.
 
If the beam is surrounded by conducting and ferromagnetic  surfaces (as in charged particle 
accelerators), then the fields around the beam change.\footnote{
  In the following, we 
distinguish between the fields created by the beam itself, i.e. own, direct fields (with the index 
``dir'') and fields from induced charges and currents (with the index ``im'').} 
To account for these changes, there is a relatively simple method developed by William Thomson 
(Lord Kelvin) \cite{Thomson_Rep_BA_1847} and described in much detail in the Maxwell treatise 
\cite{Maxwell:1873a}, referred to as the method of mirror charges and currents or the method of images.
For example, with the use of this method in \cite{Laslett:1963jn} for the first time
 1D projections  ${\rm E_y}$  and ${\rm B_x}$ of 
fields were calculated 
when 
generated by a cylindrical beam between infinitely wide parallel 
ideally conducting plates, and/or between ferromagnetic parallel poles. 
Both types of planes are parallel and symmetrical to the coordinate plane  $(x,z)$.
However, an infinite sequence of fields from  image charges and image currents 
were 
summed up only in the linear approximation in  $y$ and $\bar{y}$,
the coordinate of the field observation point on the $y$-axis, $(0,y)$
and  the beam offset from the origin, respectively.
 
As a result, due to presence of the parallel conducting and ferromagnetic plates,
additional induced fields arise near the beam
\beq
\Er_{y, \uim}(y,\bar{y})\,=
\,\frac{4 \kappa\lambda}{h^2}\epsilon_1(y+2\bar{y})\,,\qquad 
\Br_{x, \uim}(y,\bar{y})\,=
\frac{2 \kappa\lambda\beta}{cg^2}\epsilon_2(2y+ \bar{y})\,.
\label{I-3}
\eeq
Here $h$ and $g$ are half-gaps  between the conducting plates and  between the poles of  magnets, correspondingly.
The coefficients $\epsilon_1=\pi^2/48$ and $\epsilon_2=\pi^2/24$ are called the Laslett form factors for infinite  parallel plate vacuum chambers and magnetic poles, 
respectively. 
The  truncated linear approximation  \eqref{I-3} is widespread in textbooks, reference books and lectures (see, for example, \cite{Wiedemann:2007yf, Chao:1993zn, Hofmann:1992}).\footnote{We  understand 
under the complete linear approximation an expression linear in $y$, but summed up in all orders of ${\bar y}$.}
Conditions for the applicability of this linear approximation are violated if the observation 
point of the field $y$ is located far off the beam, and if the center of a beam $\bar{y}$ is far 
shifted toward the conducting surface.

The exact solution of this one dimensional problem by the method of images was presented  by the author in \cite{ Levchenko:2010pri, Levchenko:2020},
\beq 
\Er_{y, \uim}(\delta, \bar{\delta}) = \frac{4\kappa\lambda}{h}
\Lambda(\delta,\bar{\delta} )\,,\qquad
\Br_{x,\uim}(\eta,\bar{\eta}) =\frac{4\kappa\lambda\beta}{gc}
H(\eta,\bar{\eta})\,,
\label{I-4}
\eeq
where the structural functions of the fields are of the form
\begin{eqnarray}
\Lambda(\delta,\bar{\delta}) &=&
\frac{1}{2}
\Big \{ \frac{\pi}{4}\tan\Big[\frac{\pi}{4}(\delta + \bar{\delta})\Big] 
+ \frac{\pi}{4}\cot \Big[ \frac{\pi}{4}(\delta - \bar{\delta})\Big ] 
-\frac{1}{\delta - \bar{\delta}}\Big \}\,, \nonumber\\ 
H(\eta,\bar{\eta}) &=&
\frac{1}{2} \Big \{\frac{\pi}{4}
\tan\Big[\frac{\pi}{4}(\eta+\bar{\eta})\Big]
-\frac{\pi}{4}\cot\Big[\frac{\pi}{4}(\eta-\bar{\eta})\Big]
+\frac{1}{\eta-\bar{\eta}}\Big \}\,.
\label{I-5}
\end{eqnarray}
The variables, $\delta= y/h$,  $\bar{\delta}= \bar{y}/h $, $\eta=y/g$, $\bar{\eta}= \bar{y}/g$ 
correspond to the scaled  $y$ and $\bar{y}$ coordinates.
By expanding trigonometric functions into series, one can obtain both approximations (\ref{I-3}) and the full linear approximations, including generalized form factors $\epsilon_1(\bar{\delta})$,
 $\epsilon_1(\bar{\eta})$,  valid for arbitrary $\bar{\delta}$ and $\bar{\eta}$ 
\cite{Levchenko:2020}. The complete linear approximations are  important in the study of dynamics
of particles  near the beam 
and the beam itself with a significant deviation from the median plane \cite{Levchenko:2020}. 

The representation of a vacuum chamber and magnetic poles in the form of infinite parallel
plates is a very useful mathematical abstraction. 
However, in real accelerators, all components are finite although some of these 
components include elements which are structurally designed as parallel conductive and ferromagnetic  
flat surfaces. In circular accelerators such as LHC \cite{Bruning:2004}  
and the future HL-LHC \cite{HLLHC}, 
flat parallel surfaces are  parts of different types of collimators, 
 the normal conducting  separator and orbit correction dipole 
magnets.\footnote{A list of collimators for the LHC Run 2 (in 2015) includes  108 items
and shown on p. 151 of the technical design report  
``High-Luminosity Large Hadron Collider (HL-LHC)'' \cite{HLLHC}.}
As a rule, collimator jaws have a length of 600--1400 mm, 
and their width is about 90 mm,
with a jaw flatness of about 40 $\mu$m along the 1 m long active jaw surface.
Similarly, the poles of a dipole magnet 
have a length of 2000--3400 mm and  a pole width of 60 mm \cite{HLLHC}.
With transverse beam sizes as small as 200 $\mu$m, 
the representation of collimators and dipole magnets in the form of infinite  parallel 
plates is a good approximation for these  elements
and it is therefore legitimate to apply results obtained here in various applications.

To calculate the particle trajectories in the halo of the beam or 
the evolution of electrons
emitted from the jaws of a collimator, it is necessary to know the two-dimensional (2D)
distribution of  fields in the gap between the plates.\footnote{A study of
QED effects in the beam fields 
will be presented elsewhere.} 
All these problems are relevant for the HL-LHC 
under construction \cite{CERNcr19} and discussed partly in section~\ref{s6}.
In the following sections we present exact 2D 
solutions for the fields $\mathbf{E}_{\uim}$ and $\mathbf{B}_{\uim}$ obtained by the method of image charges 
and currents.

\section{2D electric field from image-charges}

\begin{figure*}[h]
\centering
\includegraphics[height=9.0cm]{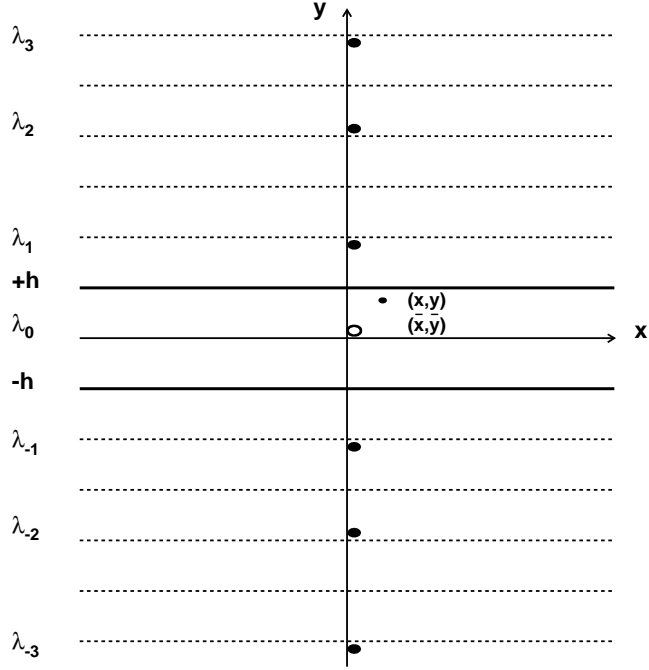}
\caption{The electric field in the point $(x,y)$ between conducting plates $y=\pm h$
is generated by the direct source-charge $\lambda_0 $ at  $(\bar{x},\bar{y})$ (the direct field)
and the successive image charges $\lambda_{\pm i}$ at locations 
$(\bar{x}, d_{\pm k})$ and $(\bar{x}, d_{\pm m})$ (see explanation in text).}
\label{fig1} 
\end{figure*}

Let us calculate the field  $\mathbf{E}_{\uim}$ in the space between perfectly conducting parallel 
infinite plates, when the particle beam moves parallel to the plates.
Let us direct the $z$-axis of the right-handed Cartesian 
coordinate system $(x, y, z)$ along the velocity 
vector of the beam particles, and also direct the  $y$-axis perpendicular to the conducting plates. 
The plates themselves will be placed at a distance $y =\pm h$ from the origin. Then the $x$-axis will 
be perpendicular to the beam and lies in the median plane between the plates.
Suppose that the constituents of the beam are positively charged. For example, it can be a beam of protons or ions.
 By construction, our problem is two-dimensional and static and, therefore, it is necessary to 
calculate the field components only in the $x$--$y$ plane. For full generality, let the beam center be 
displaced  to the point $(\bar{x},\bar{y})$, and the observation point of the field be at $(x,y)$  between conducting parallel plates. This is shown in figure~\ref{fig1}. 

The boundary condition on the $x$-component of the electric field on the surface of perfectly 
conducting plates is $E_x(x,\pm h)=0$ (equipotential surfaces) and is satisfied if image charges 
change sign from image to image. 

A test charged particle placed at a point $(x,y)$ is affected by both the direct field from 
the source-charge $\lambda_0 $ and the fields of all image charges $\lambda_{\pm i}$ 
For instance, charges\footnote{
Below, for short, instead of the term ``charge-image'', we will write 
``charge''.} 
$\lambda_1$ and  $\lambda_{-1}$ are generated by $\lambda_0 $
due to reflection in plates $+h$ and   $-h$, respectively. The  charges $\lambda_2$ and  
$\lambda_{-2}$ are generated by $\lambda_{-1}$ and $\lambda_1$ due to reflection in plates $+h$ and   
$-h$, respectively, and so on. With the help of figure~\ref{fig1}, one can easily calculate 
$y$-coordinates of all  image charges. So, for odd images, $k=1,3,5,...$,
 the differences of $y$-coordinates between $\lambda_{\pm k}$ and the observation point
$y$ are  $d_{\pm k} = 2k h \mp y_{+}$. For even images, $m=2,4,6,...$, the differences of 
$y$-coordinates between $\lambda_{\pm m}$ and the observation point $y$ are  
$d_{\pm m} = 2m h \mp y_{-}$. Here, $y_{+}=y+\bar{y}$ and $y_{-}=y-\bar{y}$.
 
The successful solution of the 1D problem was due to the fact that the distance between the field points was determined by a simple difference in coordinates, and the need to summarize a series of elementary fractions \cite{Levchenko:2020}. In the 2D problem, a distance is already specified through a power function. The transition to the  potential  greatly facilitates the solution of electrostatic problems.

In the complex $z$-plane, $z=x+i y$, we define the field by $\Er(z)=-\partial\Phi /\partial z$. 
If we set the complex  potential of the beam with a linear charge density $\lambda$  as 
$\Phi(z-\bar{z})=-2\kappa  \lambda\ln(z-\bar{z})\,$
then the complex field of such a beam is 
$\Er(z) = 2\kappa  \lambda/(z-\bar{z})$\,.
 Thus, in the components, the direct beam field is 
\beq
\Er_{\udr}(z) =
 \frac{2\kappa  \lambda}{|r-\bar{r}|}\cdot \frac{x-\bar{x}}{|r-\bar{r}|} - 
i \frac{2\kappa  \lambda}{|r-\bar{r}|}\cdot \frac{y-\bar{y}}{|r-\bar{r}|}
=\Er_{x,\udr} -i \Er_{y,\udr}\,.
\label{b3}
\eeq

Let us compose the sum of the potentials from the first-level images $\lambda_1$ 
and $\lambda_{-1}$ (figure~\ref{fig1}),
\begin{eqnarray} 
\Phi_1&=& -2\kappa  \lambda \big [(-1)\ln(z-z_1)+(-1)\ln(z-z_{-1}) \big ]=\nonumber \\
&=& 2\kappa \lambda\big \{\ln[x-\bar{x} + i(y_{+}-2h)] 
+ \ln[x-\bar{x} + i(y_{+}+2h)]\big\} = \nonumber \\
&=& 2\kappa \lambda\ln[\hat{z}_{+}^2+(2h)^2]\,,
\label{b7}
\end{eqnarray}
with $ \hat{z}_{+}=x-\bar{x}+ i y_{+}$.
For an arbitrary odd $k=2n+1$,
\beq
\Phi_k=2\kappa \lambda\ln[\hat{z}_{+}^2+(2kh)^2]
=2\kappa \lambda\ln\Big\{(2kh)^2 \Big[1+\frac{(\hat{z}_{+}/2h)^2}{(2n+1)^2}  \Big ]
\Big\}\,.
\label{b8}
\eeq
The sum of potentials from images with even $m$, $\lambda_m$ and $\lambda_{-m}$, is
\begin{eqnarray} 
\Phi_m&=& -2\kappa \lambda\big \{\ln[x-\bar{x}+ i(y_{-}-2mh)] 
+\ln[x-\bar{x}+ i(y_{-}+2mh)]\big \}= \nonumber \\
&=&-2\kappa \lambda\ln\Big\{(2mh)^2 \Big[1+\frac{(\hat{z}_{-}/2h)^2}{(2n+2)^2}  \Big ]\,,
\label{b9}
\end{eqnarray}
where $ \hat{z}_{-}=x-\bar{x}+ i y_{-}$.
The resulting potential is $\Phi_{\uim}= \sum_k \Phi_k +\sum_m \Phi_m$,
\beq
\Phi_{\uim}(\hat{z}_{+},\hat{z}_{-})=2\kappa \lambda \bigg[ 
\ln\frac{ \prod\limits_{n=0}^{\infty}(2n+1)^2}
{\prod\limits_{n=0}^{\infty}(2n+2)^2}
+\ln \prod_{n=0}^{\infty}\Big(1+\frac{(\hat{z}_{+}/2h)^2}{(2n+1)^2}  \Big ) 
-\ln \prod_{n=0}^{\infty}\Big(1+\frac{(\hat{z}_{-}/2h)^2}{(2n+2)^2}  \Big )
\bigg ]\,.
\label{b11}
\eeq
In calculating the field components, the first constant term does not contribute.

We apply now the representation of  hyperbolic functions in the form of infinite products
and transform eq. (\ref{b11}) to the following form
\beq
\Phi_{\uim}(\hat{z}_{+},\hat{z}_{-})=2\kappa \lambda \Big \{ \ln [\cosh(\hat{\delta}_{+})] -\ln [\sinh(\hat{\delta}_{-})] + \ln \hat{\delta}_{-} \Big \}\,,
\label{b13}
\eeq
where $\hat{\delta}_{+}=\pi\hat{z}_{+}/4h$ and $\hat{\delta}_{-}=\pi\hat{z}_{-}/4h$.

In an old textbook \cite{Smythe:1950}, a similar problem for the electrostatic field  was solved by the method of conformal mapping. Only the electrostatic field potential was calculated.
The expressions for the potential coincide with eq. (\ref{b13}) 
if $a=2h$ and $b=h+\bar{y}$ are used in formula (3), 
section~4.20 of \cite{Smythe:1950}, and normalised variables are not used.

We find an expression for the  complex field of electrical images by
differentiating the complex potential (\ref{b13}) with respect to  $z$, 
\beq
\Er_{\uim}(z)= \frac{2\kappa \lambda}{h} \frac{\pi}{4}\Big [\coth(\hat{\delta}_{-}) -
\tanh(\hat{\delta}_{+}) -\frac{1}{\hat{\delta}_{-}} \Big ]\,.
\label{b14}
\eeq
The real components of the electric field, $E_{x,\uim}$ and $E_{y,\uim}$, in accordance with the definition (\ref{b3}), are 
 \beq
\Er_{x,\uim}= -{\rm Re}\frac{\partial\Phi_{\uim}}{\partial z}\,, \qquad
 \Er_{y,\uim}= {\rm Im}\frac{\partial\Phi_{\uim}}{\partial z}\,.
\label{b15}
\eeq 

Let us introduce the scaled coordinates of the field observation point, $\delta_x=x/h$, $\delta_y=y/h$,
 the scaled coordinates of the beam position, $\bar{\delta}_x=\bar{x}/h$, $\bar{\delta}_y=\bar{y}/h$
and also variables $\Delta_{x}=\delta_x-\bar{\delta}_x$, $\Delta_{y-}=\delta_y-\bar{\delta}_y$, 
$\Delta_{y+}=\delta_y +\bar{\delta}_y$.
After simple but long algebraic transformations, we get the following formulas for  components of the resulting image field, 
%
\begin{eqnarray}
\Er_{x,\uim}(x,y)& = & \frac{2\kappa \lambda}{h} \frac{\pi}{8}
\frac{\sinh(\frac{\pi}{2}\Delta_{x}) \cos(\frac{\pi}{2}\delta_y) \cos(\frac{\pi}{2}\bar{\delta_y})}
{\Big [\sinh^2(\frac{\pi}{4}\Delta_{x})+\cos^2(\frac{\pi}{4}\Delta_{y+})\Big]
 \Big [\sinh^2(\frac{\pi}{4}\Delta_{x})+\sin^2(\frac{\pi}{4}\Delta_{y-})\Big]} - \nonumber \\
&-&\frac{2\kappa \lambda}{h}\frac{\Delta_{x}}{\Delta^2_{x}+\Delta_{y-}^2}\,,
\label{b4}
\end{eqnarray}
%
\begin{eqnarray}
\Er_{y,\uim}(x,y)& = & \frac{2\kappa \lambda}{h} \frac{\pi}{8}
\Big \{
\frac{\sin(\frac{\pi}{2}\Delta_{y+})}
{\sinh^2(\frac{\pi}{4}\Delta_x)+\cos^2(\frac{\pi}{4}\Delta_{y+})} 
+
\frac{\sin(\frac{\pi}{2}\Delta_{y-})}
{\sinh^2(\frac{\pi}{4}\Delta_x)+\sin^2(\frac{\pi}{4}\Delta_{y-})} 
\Big \} - \nonumber \\
&-&
\frac{2\kappa \lambda}{h}\frac{\Delta_{y-}}{\Delta_x^2+\Delta_{y-}^2}\,.
\label{b5}
\end{eqnarray}
%
To obtain the complete distribution of the electric field between the conducting plates 
and to meet the boundary condition that the electric field enters conducting surfaces
perpendicularly,
image fields $\mathbf{E}_{\uim}$ must be added to the direct field of the beam,
\beq
\Er_{x,\udr}(x,y)= \frac{2\kappa \lambda}{h}\frac{\Delta_x}{\Delta_x^2+\Delta_{y-}^2}\,,\qquad 
\Er_{y,\udr}(x,y)= \frac{2\kappa \lambda}{h}\frac{\Delta_{y-}}{\Delta_x^2+\Delta_{y-}^2}\,.
\label{b6}
\eeq
It is interesting to note some features of eqs. (\ref{b4}) and (\ref{b5}).
First, the coordinates of the field observation point enters only via the scaled variables. 
In addition, the last terms in eqs. (\ref{b4}) and (\ref{b5}) have signs opposite to those of the components of the direct field of the beam, eqs. (\ref{b6}).
Adding the components of eqs. (\ref{b4}), (\ref{b5}) and (\ref{b6}), we find the total field to be
\begin{eqnarray}
\Er_{x,\rm tot}(\delta_x,\delta_y,\bar{\delta_x},\bar{\delta_y})
&=& \frac{2\kappa \lambda}{h}\frac{\pi}{8}
\frac{\sinh[\frac{\pi}{2}(\delta_x-\bar{\delta}_x)]  }
{\Big [\sinh^2[\frac{\pi}{4}(\delta_x-\bar{\delta}_x)]+\cos^2[\frac{\pi}{4}(\delta_y+\bar{\delta_y})]\Big]}\times \nonumber \\
&\times&
\frac{\cos(\frac{\pi}{2}\delta_y) \cos(\frac{\pi}{2}\bar{\delta_y})}
{\Big [\sinh^2[\frac{\pi}{4}(\delta_x-\bar{\delta}_x)]+\sin^2[\frac{\pi}{4}(\delta_y-\bar{\delta_y})]\Big]
}\,,
\label{b16}
\end{eqnarray}
\begin{eqnarray} 
\Er_{y,\rm tot}(\delta_x,\delta_y,\bar{\delta_x},\bar{\delta_y})&=& \frac{2\kappa \lambda}{h}\frac{\pi}{8}
\bigg \{
\frac{\sin[\frac{\pi}{2}(\delta_y+\bar{\delta_y})]}
{\sinh^2[\frac{\pi}{4}(\delta_x-\bar{\delta}_x)]+\cos^2[\frac{\pi}{4}(\delta_y+\bar{\delta_y})]}+
\nonumber \\
&+&
\frac{\sin[\frac{\pi}{2}(\delta_y-\bar{\delta_y})]}
{\sinh^2[\frac{\pi}{4}(\delta_x-\bar{\delta}_x)]+\sin^2[\frac{\pi}{4}(\delta_y-\bar{\delta_y})]} 
\bigg \}\,.
\label{b17}
\end{eqnarray}
A direct verification shows that eq. (\ref{b16}) satisfies the boundary condition
$\Er_{x,\rm tot}(\delta_x,\pm 1, \bar{\delta_x},\bar{\delta_y})=0$
and in a 1D limit, $\delta_x=\bar{\delta_x}=0$ and $\Er_{x,\rm tot}=0$.
Similarly, eq. (\ref{b5}) at $\delta_x=\bar{\delta_x}=0$ goes into eq. (\ref{I-4}), with 
$\delta=\delta_y$ and $\bar{\delta}=\bar{\delta_y}$.
In particular, a beam moving in the midplane generates the field described by
$$
E_{y, \rm tot}(\delta_x,\delta_y,0,0)\,=\,\frac{2\kappa\lambda}{h}\frac{\pi}{8} 
\frac{\sin(\frac{\pi}{2}\delta_y)[1+2 \sinh^2(\frac{\pi}{4}\delta_x)]}
{[\sinh^2(\frac{\pi}{4}\delta_x)+\sin^2(\frac{\pi}{4}\delta_y)]
[\sinh^2(\frac{\pi}{4}\delta_x)+\cos^2(\frac{\pi}{4}\delta_y)]}\,.
$$
In other words, in the presence of conducting plates the electric field 
in front of the plate, $\delta_y=1$, $\delta_x=0$ is enhanced by the factor $\pi/2$.

\section{2D magnetic field from image-currents}

In the above, we have discussed electric image fields created by an
ultra-relativistic beam. Magnetic images can be treated in much the same way.
Let the ferromagnetic boundaries be represented by a pair of infinitely wide parallel
plates at $y=\pm g$. On the surface of a ferromagnetic with a very high magnetic permeability ($\mu \to \infty$), only the normal component $\mathbf{B}$ is nonzero. This boundary condition is satisfied if the image currents have the same direction as the beam current.

From eq. (\ref{I-1}), the components of $\mathbf{B}$, given that $\urv_x=\urv_y=0$ and $\urv_z=\urv$, are
\beq
\Br_x=-\beta \Er_y/c\,, \qquad  \Br_y=\beta \Er_x/c\,.
\label{c2}
\eeq
With the use of
the relationship (\ref{c2}), we apply now the method of  complex potential, as in the previous section.
After self-evident replacements in (\ref{b8}) and (\ref{b9}), we obtain an analogue of (\ref{b13}),
\beq
{\Psi}_{\uim}(\hat{\eta}_{+},\hat{\eta}_{-})=
-2\kappa \lambda (\beta/c) \Big \{ \ln [\cosh(\hat{\eta}_{+} )] 
+\ln [\sinh(\hat{\eta}_{-} )] - \ln \hat{\eta}_{-} \Big \}\,,
\label{c3}
\eeq
where $\hat{\eta}_{+}=\pi\hat{z}_+/4g$ and $\hat{\eta}_{-}=\pi\hat{z}_{-}/4g$ are normalized to the half-gap $g$ between ferromagnetic poles.
For the resulting complex magnetic induction 
from image currents, we find
 \beq
\Br_{\uim}(z)= \frac{2\kappa \lambda \beta}{g c} \frac{\pi}{4}\Big [\tanh(\hat{\eta}_{+})
+ \coth(\hat{\eta}_{-}) 
- \frac{1}{\hat{\eta}_{-}} \Big ]\,.
\label{c3-1}
\eeq

Components of the magnetic induction vector are calculated by 
$\Br_{x,\uim}= -{\rm Im}\,\partial{\Psi}_{\uim}/\partial z$,
$\Br_{y,\uim}= -{\rm Re}\,\partial{\Psi}_{\uim}/\partial z$, 
or  directly from  (\ref{c3-1}). 
In order to  obtain the complete distribution of the magnetic field between the ferromagnetic plates 
and  fulfil the boundary condition on the plates, it is necessary to add the  field $\mathbf{ B}_{\uim}$  
to the self-field of the beam. 
As previously for the electric field, adding each component $\Br_{x,\uim}$, $\Br_{y,\uim}$ 
and $\Br_{x,\udr}$, $\Br_{y,\udr}$, we find the components of the vector 
of total magnetic induction
\begin{eqnarray}
\Br_{x,\rm tot}(\eta_x,\eta_y,\bar{\eta}_x,\bar{\eta}_y)
&=&\frac{\kappa \pi\lambda \beta}{4gc} 
\bigg \{
\frac{\sin[\frac{\pi}{2}(\eta_y+\bar{\eta}_y)]}
{\sinh^2[\frac{\pi}{4}(\eta_x-\bar{\eta}_x)] + \cos^2[\frac{\pi}{4}(\eta_y+\bar{\eta}_y)]}\,-
\nonumber \\
&-&\frac{\sin[\frac{\pi}{2}(\eta_y-\bar{\eta}_y)]}
{\sinh^2[\frac{\pi}{4}(\eta_x-\bar{\eta}_x)] 
+\sin^2[\frac{\pi}{4}(\eta_y-\bar{\eta}_y)]}
\bigg \}\,,
\label{c8}
\end{eqnarray}
\begin{eqnarray} 
\Br_{y,\rm tot}(\eta_x,\eta_y,\bar{\eta}_x,\bar{\eta}_y)&=&
\frac{\kappa \pi\lambda \beta}{4gc} 
\bigg \{
\frac{\sinh[\frac{\pi}{2}(\eta_x-\bar{\eta}_x)]}
{\sinh^2[\frac{\pi}{4}(\eta_x-\bar{\eta}_x)] +\cos^2[\frac{\pi}{4}(\eta_y+\bar{\eta}_y)]} +
\nonumber \\
&+&\frac{\sinh[\frac{\pi}{2}(\eta_x-\bar{\eta}_x)]}
{\sinh^2[\frac{\pi}{4}(\eta_x-\bar{\eta}_x)] 
+\sin^2[\frac{\pi}{4}(\eta_y-\bar{\eta}_y)]}
\bigg \}\,,
\label{c9}
\end{eqnarray}
where $\eta_x= x/g$, $\bar{\eta}_x= \bar{x}/g$, 
$\eta_y= y/g$,  $\bar{\eta}_y= \bar{y}/g$.

Directly from (\ref{c8}) it is difficult to see that at $\eta_y=\pm 1$ the boundary condition 
$\Br_{x,\rm tot}(\eta_x,\pm 1,\bar{\eta}_x,\bar{\eta}_y)=0$
is satisfied. However, it can be verified that 
$\Br_{x,\rm tot}\sim\cos(\pi\eta_y/2)$, 
and thus the specified boundary condition is indeed satisfied.

\section{Surface charge density}

The distribution of the charge density $\sigma$ induced by a beam on the surface of a conductor is determined by the normal component $\Er_n$ of the electric field at a given point on the surface.
For our problem, $\Er_n= -\Er_{y,\rm tot}(\delta_x,\pm 1,\bar{\delta_x},\bar{\delta_y})$.
The charge distributions  $\sigma$ are different on the upper, $\delta_y=+1$,
 and lower, $\delta_y=-1$, plates.
Thus, according to the Gauss-Ostrogradsky theorem,
\beq
\sigma_{\pm 1}(\delta_x,\bar{\delta_x},\bar{\delta_y})=- \epsilon_0 \Er_{y,\rm tot}(\delta_x,\pm 1,\bar{\delta_x},\bar{\delta_y})\,,
\label{c11}
\eeq
where $\Er_{y,\rm tot}$ is given by eq.~(\ref{b17}).

\section{Linear approximations for image fields}

An interesting mathematical feature of the exact functions 
$\Lambda(\delta,\bar{\delta})$ and $H(\eta,\bar{\eta})$ from (\ref{I-4}), was noted in 
\cite{Levchenko:2020}.
This is a kind of permutation symmetry that connects the components of the fields
$\Er_{y, \uim}$ and $\Br_{x,\uim}$.
This functional symmetry is obvious from (\ref{I-5}): 
$H(\eta,\bar{\eta})\,=\,\Lambda (\bar{\eta},\eta )$.
In 2D solutions, (\ref{b17}) and (\ref{c8}), the functional parts of formulas retain this symmetry, 
as is directly verified. This symmetry is also present in (\ref{I-3}).

It is quite time-consuming to derive  the linear approximation for the components of 
$\mathbf{ E}_{\uim}(x,y)$ directly from eqs. (\ref{b16}), (\ref{b17}).
The result is achieved faster with the use of the complex form (\ref{b14}).
Let us decompose the function of two variables $\Er_{\uim}(z)$ in a series near 
the point $(\bar{x}, \bar{y})$ and keep the terms linear in $x-\bar{x}$  and $y-\bar{y}$.
However, when decomposing $\coth(\hat{\delta}_{-})$, we have to account
the presence of a singular term,  $\coth(\hat{\delta}_-) \approx \hat{\delta}_{-}^{-1} + \hat{\delta}_{-}/3 + \cdots$.
Thus,
\beq
\Er_{\uim}(z)\approx -\frac{4\kappa\lambda}{h}
\Big \{ \epsilon_1(\bar{\delta_y})[(\delta_x - \bar{\delta}_x) + i(\delta_y - \bar{\delta}_y) ]
+ i\frac{\pi}{8}\tan \Big(\frac{\pi}{2}\bar{\delta}_y \Big)
\Big \}\,,
\label{s5-1}
\eeq
or, in accordance with (\ref{b3}),
\beq
\Er_{x,\uim}\approx -\frac{4\kappa\lambda}{h}\epsilon_1(\bar{\delta_y})(\delta_x - \bar{\delta}_x)\,,
\qquad 
\Er_{y,\uim}\approx \frac{4\kappa\lambda}{h}
\Big [\frac{\pi}{8}\tan \Big(\frac{\pi}{2}\bar{\delta}_y \Big) + 
\epsilon_1(\bar{\delta_y})(\delta_y - \bar{\delta}_y) \Big ]\,.
\label{s5-2}
\eeq
Here we have introduced  \cite{Levchenko:2020} a generalization  of the Laslett electric image coefficient $\epsilon_1$  in the case of an arbitrary beam offset $\bar{\delta}_y$,
\beq
\epsilon_1(\bar{\delta}_y)\,=\,\frac{\pi^2}{32}\Big [\frac{1}{\cos^2(\frac{\pi}{2}\bar{\delta}_y )} 
- \frac{1}{3} \Big ],\qquad \epsilon_1(0)\,=\,\frac{\pi^2}{48}\,.\nonumber\\
\eeq
Thus, the vertical component of the electric field in the vicinity of the plate, 
$|\bar{\delta}_y| \sim 1$,  $|\delta_y-\bar{\delta}_y| \ll 1$  is given by
\beq
E_{y,\rm tot}(0, \delta_y, 0, \bar{\delta}_y) \approx \frac{2 \kappa\lambda(z,t)}{h}\Big 
[\frac{1}{\delta_y - \bar{\delta}_y}
+  \frac{\pi}{4}\frac{1}{\cos (\frac{\pi}{2}\bar{\delta}_y)} 
+ \frac{\pi^2}{16} \frac{(\delta_y-\bar{\delta}_y)}{\cos^2(\frac{\pi}{2}\bar{\delta}_y )}  \Big ]. 
\label{s4-9}
\eeq
Here we only retained  the main singular contributions. Equation (\ref{s4-9}) 
tell us that with an increase of $\bar{\delta}_y$ the field strength
near the plate  significantly increase due to images and  under the influence of the Lorentz force
the beam is attracted by the conducting plate.

\begin{figure*}[t]
\centering
\begin{minipage}[c]{0.49\textwidth}
\includegraphics[width=1.0\textwidth]{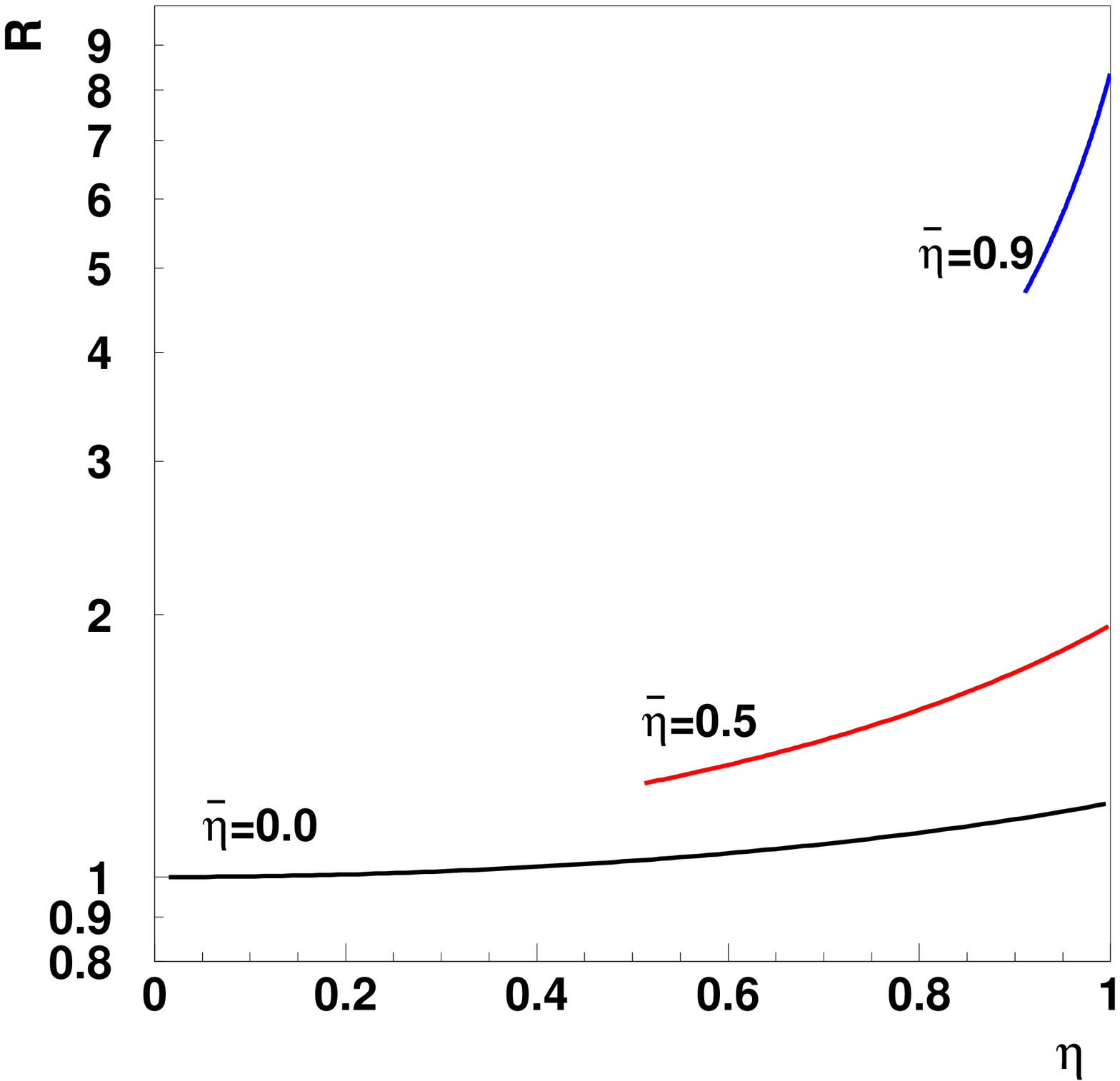}
\end{minipage}%
\begin{minipage}[c]{0.49\textwidth}
\centering
\includegraphics[width=1.0\textwidth]{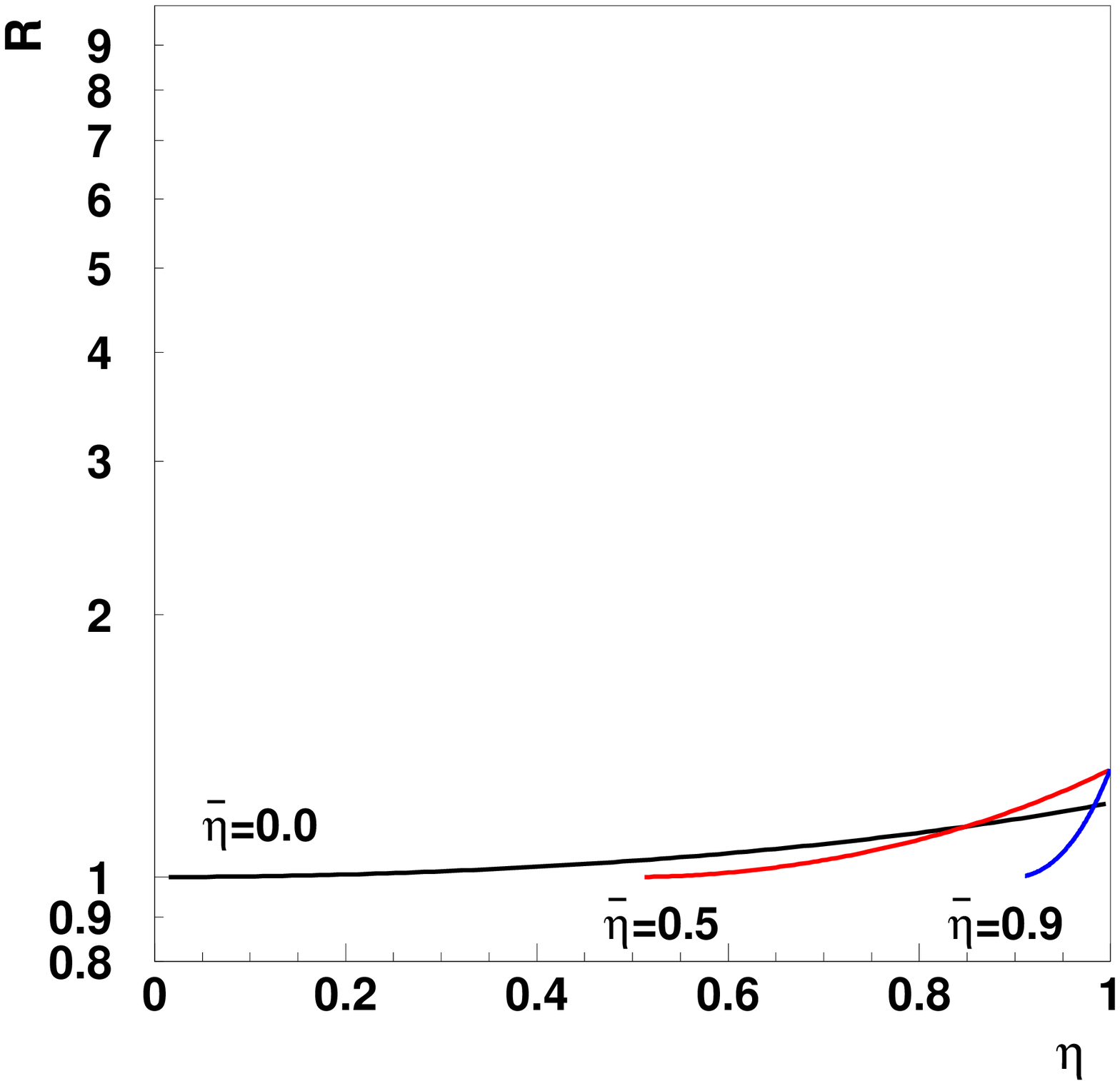}
\end{minipage}
\caption{Comparison of linear approximations for the magnetic induction of image fields   with the exact solution  at three values of the beam offset $\bar{\eta}$:
Left: The ratio of $\Br_{x,\uim}$ from  eq. (\ref{c8}) to  
$\Br_{x,\uim}$ from eq.~(\ref{I-3}); 
Right: The ratio of  $\Br_{x,\uim}$ from eq.~(\ref{c8}) to  $\Br_{x,\uim}$ from eq.~(\ref{s5-4}).}
\label{fig2}
\end{figure*}

Usually \cite{Hofmann:1992, Wiedemann:2007yf}, 
after solving a 1D problem and finding $\Er_{y,\uim}$ of the form (\ref{I-3}),
 to find the second component of the field, $\Er_{x,\uim}$, 
one use the following condition that the components of the image fields must satisfy, 
\beq
 \mathbf{ \nabla} \mathbf{ E}_{\uim} = \frac{\partial \Er_{x,\uim}}{\partial x} + \frac{\partial \Er_{y,\uim}}{\partial y} = 0\,.
\label{s5-3}
\eeq  
It is easy to verify that the field projections (\ref{s5-2})  satisfy the condition 
(\ref{s5-3}).

Now we calculate in a linear approximation the components of the magnetic induction vector.
For this, we also use the complex representation (\ref{c3-1}).
Repeating the expansion of the hyperbolic functions in a series, as in the calculation of $\Er_{\uim}$, we obtain
\beq
\Br_{x,\uim}\approx \frac{4\kappa\lambda \beta}{g c}
\Big [\frac{\pi}{8}\tan \Big(\frac{\pi}{2}\bar{\eta}_y \Big) + 
\epsilon_2(\bar{\eta_y})(\eta_y - \bar{\eta}_y) \Big ]\,,
\qquad 
\Br_{y,\uim}\approx \frac{4\kappa\lambda \beta}{g c}\epsilon_2(\bar{\eta_y})(\eta_x - \bar{\eta}_x)\,.
\label{s5-4}
\eeq
Here we have introduced \cite{Levchenko:2020, Zotter:1975} a generalization of the Laslett 
form factor for infinite parallel plate magnet poles $\epsilon_2$ in
the case of an arbitrary offset $\bar{\eta}_y$,
\beq
\epsilon_2(\bar{\eta}_y)\,=\,\frac{\pi^2}{32}\Big [\frac{1}{\cos^2(\frac{\pi}{2}\bar{\eta}_y )} 
+ \frac{1}{3} \Big ],\qquad \epsilon_2(0)\,=\,\frac{\pi^2}{24}\,.\nonumber\\
\eeq

In order to illustrate the accuracy of the
 linear approximations  (\ref{I-3}) and (\ref{s5-4})
we construct  ratios, R, of the exact solution $\Br_{x,\uim}$ from eq.~(\ref{c8}) to  $\Br_{x,\uim}$
 from eq.~(\ref{I-3}) and to $\Br_{x,\uim}$ from eq.~(\ref{s5-4}). 
Figure~\ref{fig2} shows  these ratios at three values of $\bar{\eta}_y$.
We conclude that in contrast to eq.~(\ref{I-3}), the linear approximations (\ref{s5-4})
properly take into account the dependence on  $\bar{\eta}_y$ but the range
of applicability of the modified linear approximation shrinks with an increase of $\bar{\eta}_y$. 

In conclusion, it should be noted that the condition (\ref{s5-3}) is not a universal way to recover the missing field component. Indeed, if one starts with ``a wrong component'', say, 
only the component $\Er_{x,\uim}$ is known, then it is possible to restore $\Er_{y,\uim}$
 only up to an unknown constant C. There are no additional conditions to restore this constant in the form ${\rm C}= \frac{\pi}{8}\tan \Big(\frac{\pi}{2}\bar{\delta}_y \Big)$.

Another limitation of the applicability of the linear approximation near the conductor surface,
both (\ref{I-3}) and improved (\ref{s5-2}), is the violation of the boundary condition 
$\Er_{x,\rm tot}= \Er_{x,\uim} + \Er_{x,\udr}=0$ at $\delta_y = \pm 1$.
The same applies to the component $\Br_{x,\uim}$ (\ref{s5-4}):
$\Br_{x,\rm tot}\ne 0$ at $\eta_y = \pm 1$.
The permutation symmetry mentioned at the beginning of the section for the exact expressions of 
the components $\Er_{y,\uim}$ and $\Br_{x,\uim}$, is lost in the linear approximations 
(\ref{s5-2}) and (\ref{s5-4}).

\section{\label{s6}Field emission  from jaws of a  HL-LHC collimator}

The transverse beam profile in an accelerator is formed not only by the magnetic system, but also 
 by  collimators, absorbing the halo of the beam, and protecting other parts of the accelerator 
in case of beam instability. 
As noted in section 1, the HL-LHC beam diameter, $2\sigma_x$, is much smaller than the width of
the collimator jaws and less than the working distance between the plates.
The surfaces of  collimators are made of electrically conductive alloys, 
resistant to high temperature and a mechanical stress.
An ultra-relativistic beam of charged particles is the source of a strong electric field. It is in the  collimator that the beam is closest to the electrically conductive surfaces.
As a result of these factors, the electron field emission occurs from  surfaces of 
a collimator \cite{Levchenko:2006}.
Once in the channel of the accelerator, these electrons generate a whole chain of processes that  
affect the stability of the high-energy beam. 
Accelerated by the electric field of the beam,  electrons begin to collide with the walls of the vacuum 
chamber, knock out additional electrons and emit bremsstrahlung photons, thereby leading to cascade 
multiplication of the number of electrons \cite{ecloud:2004}. 
An increasing number of electrons leads to  partial neutralization of the beam space charge, 
violating acceleration dynamics and enhancing the effects of instability.
 A summary of the Fowler-Nordheim theory of field emission is presented in appendix  \ref{appA}.  
The issues raised here are discussed in more detail
in the review \cite{Levchenko:2006}.

In modern accelerators the beam split into bunches. Here we consider bunches shaped as a 
cylinder of length $L_b$ with a circular  cross section. The coordinate $z$-axis is along the 
bunch velocity vector. 
The space-time distribution of the electric field around an ultra-relativistic 
circular bunch with a uniform particle density is described by a step-like form 
\cite{Levchenko:2020}:
\beq
 \mathbf{E}(r,z,t)\,=\,\frac{2\kappa\lambda}{r}\frac{\mathbf{r}}{r}
\Big [\theta (z-\beta ct) -\theta (z-\beta ct -L_b) \Big ]\,.
\label{s6_1}
\eeq 
This field contains only the radial component and in the functional sense at every moment
 coincides with the field of a continuous beam (\ref{I-1}).
For this reason, formulae derived in sections~2 and 3 
 for  fields of a continuous beam between parallel plates are also applicable for calculating fields of an ultra-relativistic bunch \cite{Levchenko:2020}.

We apply eq.~(\ref{b17}) 
 and  formulae from 
appendix \ref{appA} for evaluating 
the effective field emission current and the electron emission intensity 
from the jaw surface in one of  the HL-LHC collimators.
The emission current density $\mathbf{J}$ depends not only on the magnitude of the electric field on 
the surface, but also on the morphological features of the conductor surface that can locally 
significantly strengthen the applied electric field. The deviation of the real surface (with microscopic protrusions, apexes, scratches) from a perfectly smooth surface is characterized by the
field enhancement factor $\beta_{_{FN}}\geq 1$. For a particular surface, 
an average $\bar{\beta}_{_{FN}}$ is determined by measurements. 
Collimators with working surfaces made of different materials (CFC, MoGr, Mo) are under study for the HL-LHC \cite{HLLHC}. Based on the surface photos of collimator prototypes \cite{HLLHC}, 
 we consider that
the field enhancement $\bar{\beta}_{_{FN}}$ can reach a value of several tens of units.

\begin{figure*}[t]
\centering
\begin{minipage}[c]{0.5\textwidth}
\includegraphics[width=1.0\textwidth]{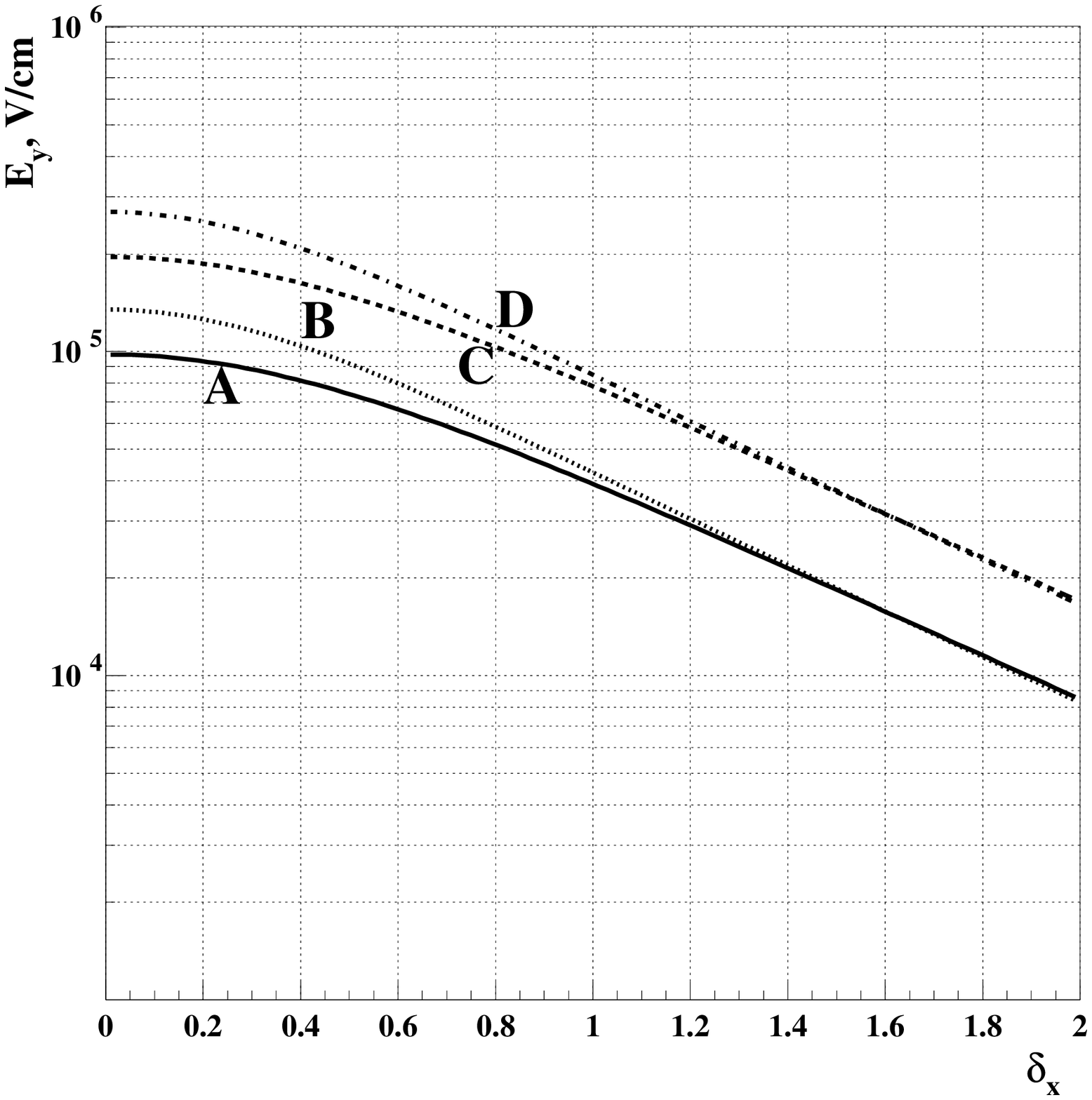}
\end{minipage}%
\begin{minipage}[c]{0.5\textwidth}
\centering
\includegraphics[width=1.0\textwidth]{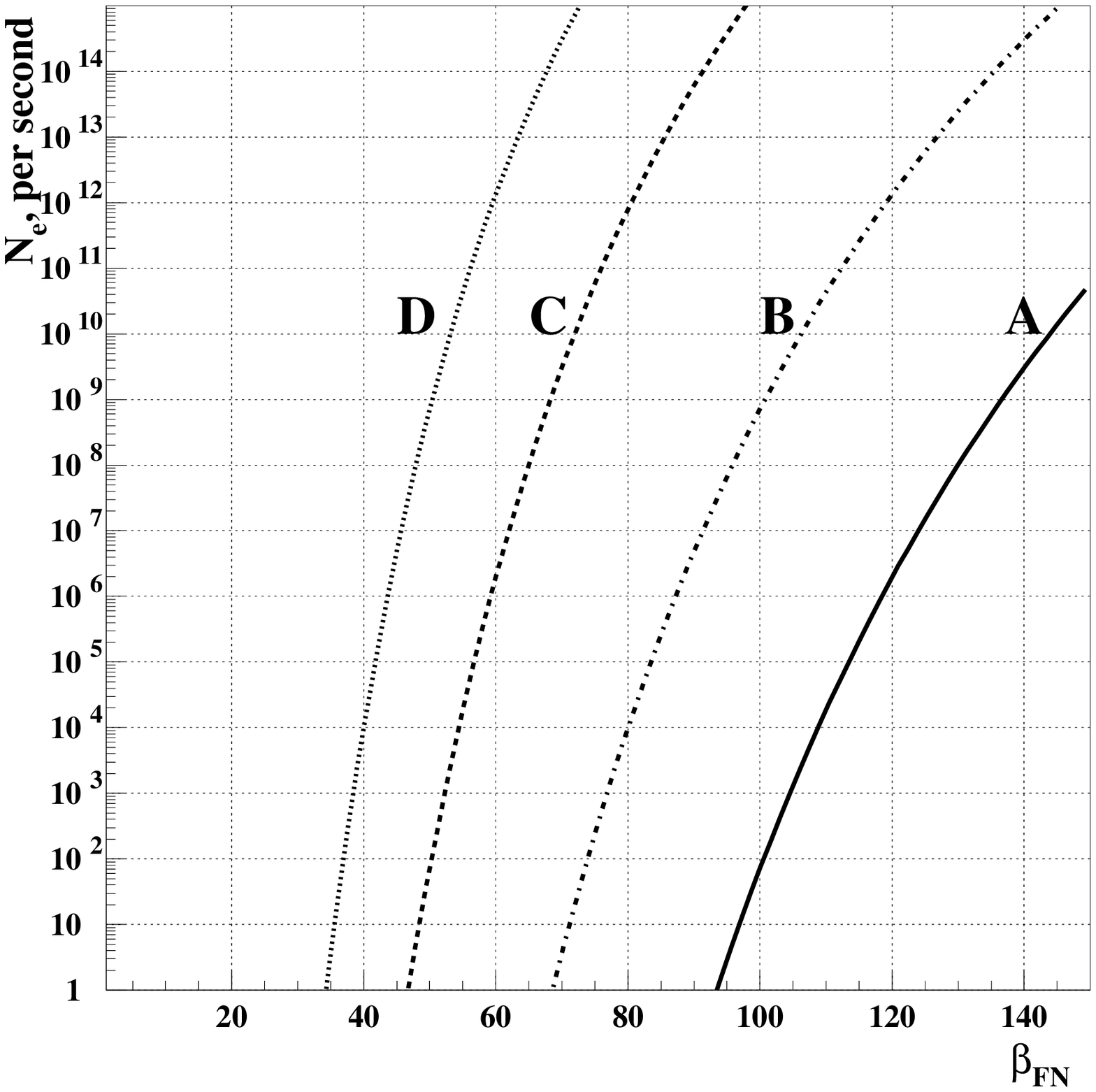}
\end{minipage}
\caption{Left: The electric field strength profile at a collimator jaw surface 
as a function of the scaled $x$-coordinate. 
Right: The electron emission intensity from a molybdenum-coated collimator
as a function of $\beta_{_{FN}}$.
Calculations are performed for
  different scenarios of jaw opening $h$
and the beam displacement from the midplane $\bar{\delta_y}$: 
A - $2h$= 1 mm, $\bar{\delta_y}=0$; B - $2h$=  1 mm, 
$\bar{\delta_y}=0.2$; C - $2h$=  0.5 mm, $\bar{\delta_y}=0.0$;
D - $2h$=  0.5 mm, $\bar{\delta_y}=0.2$. 
Other parameters are as follows \cite{HLLHC}: number protons per bunch $N_p=2.2\cdot 10^{11}$,
$L_b=\sqrt{2\pi}\sigma_z=20.3$ cm, with $\sigma_z$ denoting the r.m.s. bunch length,
$\varphi_{_{Mo}}=4.27$ eV, $L_j=100$ cm, $T=327$ K.
}
\label{fig3} 
\end{figure*}

For  numerical calculations 
we use  the HL-LHC proton beam  parameters (\cite{HLLHC}, table 2-1) and parameters of a
TCSMP/TCTPM collimator with a molybdenum surface (\cite{HLLHC}, tables 5-3, 5-4).
Figure~\ref{fig3}:Left shows the distribution of electric field intensity on a jaw surface.
The number of electrons  per second emitted from the collimator jaws is
\beq
N_e(\beta_{_{FN}})= n_b N_r\sum \int \int \int \frac{1}{e}\mathbf{J}(\beta_{_{FN}}\Er_y) \Theta (z,t)\ud x \ud z \ud t\,.
\label{s6_2}
\eeq
 Here,  the integration is carried out over  the emission area $ S(t)$   
activated  on the jaw of  length  $L_j$ by the field pulse  from a bunch of  length  $L_b$.
The function $\Theta (z,t)$ is the step-like part of eq.~(\ref{s6_1}).
The emission is summed from both jaws.
The duration of the field pulse is $\Delta t = L_j/c$. 
It is necessary to take into account that in the HL-LHC 
proton beam  there will be $n_b$=2748 bunches and per second the beam makes 
$N_r$=11245 revolutions. Appendix \ref{appB} gives details of calculating $N_e$ by the formula 
\eqref{s6_2}.

Figure~\ref{fig3}:Right shows the results of calculating $N_e(\beta_{_{FN}})$ for different 
values of $h$ and $\bar{\delta_y}$.
So, in the scenario $\mathrm{C}$, at $\beta_{_{FN}}\!\sim$80
the number of electrons emitted per second already exceeds the number of protons in one bunch,
$N_p=2.2\cdot 10^{11}$.
And with a certain shift of the beam from the 
midplane (scenario $\mathrm{D}$, $\bar{\delta_y}=0.2$), 
the number of emitted electrons increases by four orders of magnitude 
and exceeds the total number of protons in the beam, the order of $6.0\cdot 10^{14}$.
 It should be noted that our estimations does not take into account contributions 
from other collimators, the dynamical multiplication of the number of electrons 
during rescattering, as well as electron absorption in other parts of the accelerator, for instance, in a special beam screen.

\section{Summary}

We analyzed the problem of summing up the fields of the images generated by a charged beam between
infinitely wide parallel conducting plates and/or ferromagnetic poles. 
The new exact 2D solutions for resulting electric and magnetic  fields 
are represented in terms of elementary trigonometric and hyperbolic functions.
The  expressions for modified fields are applied to develop 
 improved linear approximations for image fields and to generalize the Laslett image coefficients
to the case of an arbitrary beam offset.
We apply the solution for the electric field between parallel conducting plates
 to calculate the surface distribution of the induced electric charge and
for evaluating the electron emission intensity from a molybdenum-coated collimator
of the future HL-LHC.
To conclude,  we note the need  ``to certify'' the jaw surface of all collimators by the value of 
$\bar{\beta}_{_{FN}}$ and  simulate the particle beam wiring throughout the accelerator  
 to determine optimal working distances between  jaws from minimizing
the  field emission.

\acknowledgments
The author is grateful to 
M. Wing   for reading the manuscript, questions and comments 
and E.B. Oborneva for discussions.

\appendix
\renewcommand{\theequation}{A.\arabic{equation}}
\renewcommand{\thefigure}{A-\arabic{figure}}
\renewcommand{\thetable}{A-\arabic{table}}
\setcounter{equation}{0}  

\section{\label{appA}Field emission in a strong electric field }

This summary of the Fowler-Nordheim theory of field emission is based on 
the review \cite{Levchenko:2006} and refs. \cite{FN,Nord28,Nord29}, \cite{elinson:1958, fursey:2005, SB,MG,Chr1,GM,Sh,Gomer,Mod}.
\subsection{Fowler-Nordheim theory }

In the framework of  Fowler-Nordheim (F-N) theory,  the current density of field 
emission of electrons from a metal can be written in the following form 
%
\beq
\mathbf{J}_{FN}\,=\,e\int n(\Ec_y )D(\Ec_y, \Er_y)d\Ec_y\,,
\label{3.1}
\eeq 
where $D(\Ec_y,\Er_y)$ is the penetration coefficient and $n(\Ec_y )$ is the number
of electrons at the energy $\Ec_y$ incident in the $y$-direction on the surface barrier 
from inside of the metal.

An electron outside a metal is attracted to the metal  as a result of the 
charge it induces on the surface (image force).
In the externally applied accelerating electric field $\Er_y$, the potential energy of the
electron is
\beq
V(y)\,=\,-\frac{e^2}{4y}\,-\,e\Er_y y\,,\ \ \ \ {\rm when} \ \ \ y>0\,,
\label{3.2}
\eeq 
where  $y$ denotes the distance from the surface
and the first term accounts for the image potential.\footnote{In this section we  adopt the Gaussian CGS system.} 
With use of 
the potential energy (\ref{3.2}) and the Fermi energy distribution of electrons 
in the conduction band,  one finds  
 that
\beq
\mathbf{J}_{_{FN}}(\Er_y)\,=\, A \frac{{\Er_y}^2}{\varphi\cdot t^2 (\tau)}\exp \Big\{-B\frac{\ 
\varphi ^{3/2}}{\Er_y}\vartheta (\tau) \Big\}\,,
\label{eq:n1}
\eeq 
where $\mathbf{J}$ is the current density  in A/cm$^2$, $\Er_y$ is electric field  on the  surface 
in V/cm, and  $\varphi$ is the work function in eV.  The field-independent constants $A$ and  $B$ 
 and the variable  $\tau$ are 
\begin{eqnarray}
A&=&\frac{e^3}{8\pi h}\,=\, 1.5414\cdot 10^{-6}\,, \qquad   
B=\frac{8\pi\sqrt{2m}}{3eh}\,=\,6.8309\cdot 10^7\,, \nonumber \\
 \tau &=& \frac{\sqrt{e^3 \Er_y}}{\varphi} = 3.7947\cdot 10^{-4} 
\frac{\sqrt{\Er_y}}{\varphi}\,,
\label{eq:n2*}
\end{eqnarray}
where $-e$ is the charge on the electron, $m$ is the electron mass
and $h$ is  Planck's constant. The numerical values of $A$ and $B$  correspond 
to  values of the physical constants   \cite{PDG}.  
We note that under field emission conditions, 0$<\tau \le$1.

The Nordheim function $\vartheta (\tau)$ takes into account  a lowering of the potential 
barrier due to the image potential (the Schottky effect) and  its distinction from an 
idealized triangular shape. 
The function $t(\tau)$ in the denominator of eq.~(\ref{eq:n1}) is defined as
\beq
t(\tau)\,=\,\vartheta (\tau)-(2\tau/3)(d\vartheta /d\tau)\,.
\label{eq:n5}
\eeq  
The function $\vartheta (\tau)$ varies from $\vartheta (0)=1$ to $\vartheta(1)=0$
 with the increase in field strength, however  $t(\tau)$  is quite close to unity 
at all values of $\tau$. 

For  a typical metallic $\varphi$ of 4.5 eV, fields of the order of 10$^7$ V/cm are needed to
have measurable emission currents. In considering magnitudes, one must always keep in mind
the rapid variation of the exponential function. For instance, an increase in $\Er$ of only a
factor of two from $1\times 10^7$ to $2\times 10^7$ V/cm increases the current density by
15 orders of magnitude (from $10^{-18}$ to $10^{-3}$A/cm$^2$)~! 

At a field strength of the order of
$\Er_{cr}=\varphi^2/e^3=6.945\cdot 10^{6}\cdot \varphi^2$ V/cm, 
the height of the potential barrier  vanishes  and $\vartheta (1)=0$. 
For instance, for copper $\varphi_{_{Cu}}=4.65$ eV giving 
$\Er_{cr}(Cu)=1.5\cdot 10^{8}$ V/cm,
and  similarly for graphite, $\varphi_{gr}=4.6$ eV,  
$\Er_{cr}(gr)=1.47\cdot 10^{8}$ V/cm.
At this field level one would expect the
orderly bound states characteristic of the solid to lose their integrity.

For a long time only tabulated values of $\vartheta (\tau)$ and $t(\tau)$
\cite{BKH}  were used in calculations, see \cite{MG,Chr1,GM,Sh,Gomer,Mod}. 
Several  parameterizations of  functions  
$\vartheta (\tau)$  and $t(\tau)$  are proposed in  ref. \cite{bbl1}.

The theory of field emission from metals has been subjected to fairly extensive
verification. A variety of methods have been employed over many
years for the experimental measurements of the emission current as a function
of  the field strength, the  work function and the energy distribution of  
the emitted electrons \cite{GM,Sh,Gomer,Mod}.
The F-N theory (\ref{eq:n1}) of electron emission from plane
and uniform metal surfaces (single-crystal plane) at $T\approx 0$
may therefore be considered well 
established experimentally as  well as on theoretical grounds.

\subsection{Temperature dependence}

The main equation (\ref{eq:n1}) of the  F-N theory was derived for an idealized
metal in the framework of the Sommerfeld model, with an ideally flat  surface and  
at a very low temperature, $T\approx 0$.
The temperature dependence of the field emission current (FEC) is completely 
connected with the change of the spectrum  of electrons in the metal with an increase in $T$.
Therefore, at non-zero temperatures the F-N theory must be modified to take into account
the thermal excitation of electrons above the Fermi level.
For the so-called  extended field emission region,  Murphy and  Good \cite{MG}
obtained the following elegant equation
\beq
\mathbf{J}_{_{FN}}(\Er_y,T)\,=\,\frac{\pi \omega}{\sin \pi \omega}\mathbf{J}_{_{FN}}(\Er_y,0)\,,
\label{s2.3.1}
\eeq 
which accounts for  the temperature dependence of the FEC.
 Here $\omega = k_{_B}T/k_{_B}T_{_0}$ and 
\beq
k_{_B}T_{_0}\,=\, \frac{2}{3}\frac{\Er_y}{B t(\tau)\sqrt{\varphi}}\,,
\label{s2.to}
\eeq 
where $k_{_B}$ is  Boltzmann's constant, and $T$ is the absolute temperature in K. 
It can be shown \cite{MG} that eq. (\ref{s2.3.1}) is a valid approximation when the
following two conditions are satisfied:
\beq
\omega <  \Big [ 1+\frac{1}{\Gamma_1}\Big ]^{-1}\,, \ \ \ \Gamma_1 =\frac{\varphi (1-\tau)}{k_{_B}T_{_0}}
-\frac{2}{\pi}\Big (\frac{2}{\tau}\Big )^{1/2}t(\tau)\,,
\label{s2.g1}
\eeq 
and
\beq
\omega  < \Big [ 1+\frac{1}{\Gamma_2}\Big ]^{-1}\,, \ \ \ \Gamma_2 \simeq
\Big (\frac{2\varphi }{k_{_B}T_{_0} t(\tau)}\Big )^{1/2}\,.
\label{s2.g2}
\eeq 
At very low temperatures, when $\pi\omega$ is small, eq.  (\ref{s2.3.1}) reduces 
to eq. (\ref{eq:n1}).
By expanding $\sin \pi \omega$ in a series, one gets  for  practical use the
formula
\beq
 \mathbf{J}_{_{FN}}(\Er_y,T)/\mathbf{J}_{_{FN}}(\Er_y,0)=1+1.40\cdot10^8(\varphi/\Er_y^2)T^2\,.
\label{s2.3.2}
\eeq 
It is easy to estimate using eq. (\ref{s2.3.2}) that for $\varphi=$4.5~eV  
at room temperature, $T=300$K, and the field strength 
$\Er_y=1\times 10^{7}$ V/cm and $2\times 10^{7}$ V/cm,
 the temperature factor in  (\ref{s2.3.1})  equals 1.57 and 1.14, respectively.
Thus, the temperature factor appear to be a sizeable correction.


\renewcommand{\theequation}{B.\arabic{equation}}
\renewcommand{\thefigure}{B-\arabic{figure}}
\renewcommand{\thetable}{B-\arabic{table}}
\setcounter{equation}{0}  
\setcounter{figure}{0}
\section{\label{appB}The size of emission area}
\begin{figure*}[t]
\centering
\includegraphics[height=7.0cm]{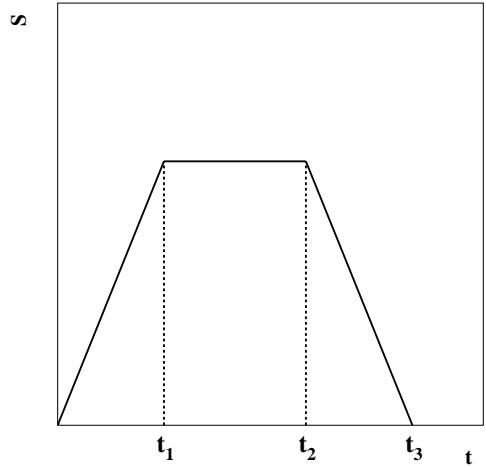}
\caption{
A schematic view of  variation in time of the emission area on the jaw surface 
covered by the electric field of a bunch (in arbitrary units).
}
\label{fig4} 
\end{figure*}

Here we derive the formula for calculating $N_e$, see eq.~(\ref{s6_1}).
Figure~\ref{fig4} shows schematically how the emission area varies over time as a bunch passes through the collimator.
Because of a step-like form of the field, the electric field of the bunch covers only a part of the jaw surface. For that reason,
\beq
 \ud x\int \Theta (z,t) \ud z \ud t =  \int S(t)\ud t\,,
\eeq
where $\Theta (z,t)=\Big [\theta (z-\beta ct) -\theta (z-\beta ct -L_b) \Big ]\,.$
There are three stages: $[0,t_1]$:\,$S(t)= \delta s\cdot t/t_1$; 
$[t_1,t_2]$:\,$S(t)=\delta s$; $[t_2,t_3]$:\,$S(t)=\delta s\cdot [1-(t-t_2)/(t_3-t_2)]$\,,
with $\delta s =\ud x\cdot L_b$. 
Therefore, 
$\int S(t)\ud t = \delta s\cdot t_2$, with $t_2=L_j/c$.
With these substitutions, eq.~(\ref{s6_1}) finally becomes
\beq
N_e(\beta_{_{FN}})= n_b N_r\frac{L_b L_j}{c}
\sum \frac{2}{e}\int_0^{\infty}\mathbf{J}(\beta_{_{FN}} E_y(x,\bar{\delta_y})) \ud x\,.
\eeq 
%


\end{document}